\documentclass[amsmath,amssymb,preprint,superscriptaddress]{revtex4}

\usepackage{graphicx}
\usepackage{acronym}

\let\phi=\varphi
\let\a=\alpha \let\b=\beta \let\g=\gamma \let\d=\delta
   \let\k=\kappa
\let\l=\lambda \let\m=\mu   
\let\s=\sigma  \let\f=\varphi 
   \let\G=\Gamma
\let\D=\Delta \let\L=\Lambda

 \def\VV{{\cal V}}

\def\DD{{\cal D}}

\def\to{\rightarrow} \def\la{\left\langle} \def\ra{\right\rangle}

\newcommand{\wh}{\widehat} 

\newcommand{\eee}{\mathrm{e}}
\newcommand{\beq}{\begin{equation}}
\newcommand{\eeq}{\end{equation}}
\newcommand{\ba}{\begin{align}}
\newcommand{\ea}{\end{align}}

\def\de{\mathrm d}

\acrodef{RFOT}{random first-order theory}
\begin{document}

\title{Breakdown of Elasticity in Amorphous Solids}

\author{Giulio Biroli} 
\affiliation{IPhT, Universit\'e Paris Saclay, CEA, CNRS, F-91191 Gif-sur-Yvette Cedex, France}
\affiliation{LPS, Ecole Normale Sup\'erieure, 24 rue Lhomond, 75231 Paris Cedex 05 - France.}
  
\author{Pierfrancesco Urbani}
\affiliation{IPhT, Universit\'e Paris Saclay, CEA, CNRS, F-91191 Gif-sur-Yvette Cedex, France}

\maketitle

{\bf  What characterises a solid is its way to respond to external stresses. Ordered solids, such crystals, display an elastic regime followed by a plastic one, both well understood microscopically in terms of lattice distortion and dislocations. For amorphous solids the situation is instead less clear, and the microscopic understanding of the response to deformation and stress is a very active research topic. Several studies have revealed that even in the elastic regime the response is very jerky at low temperature, resembling very much the one of disordered magnetic materials. Here we show that in a very large class of amorphous solids this behaviour emerges by decreasing the temperature as a phase transition where standard elastic behaviour breaks down. At the transition all non-linear elastic moduli diverge and standard elasticity theory does not hold anymore. Below the transition the response to deformation becomes history and time-dependent.}\\
 Our work connects two different lines of research. The first focuses on the behaviour of amorphous solids at low temperature. With the aim of understanding the response of glasses to deformations, there have been 
extensive numerical studies of stress versus strain curves obtained by quenching model systems at zero temperature. One of the main outcome is that 
the increase of the stress is punctuated by sudden drops related to avalanche-like rearrangements both before and after the yielding point \cite{BL11,ML06, LC09,KLP10, LLRW14}.
%TO CITE in order of time (older first), and first of all Barrat since it's a review
%BarratLemaitre Barrat, Jean-Louis, and Ana'l Lemaöõtre. "Heterogeneities in amorphous systems under shear." Dynamical %Heterogeneities in Glasses, Colloids, and Granular Media 150 (2011): 264.
%LemaitreCaroli Phys Rev E Stat Nonlin Soft Matter Phys 103(6):065501.
%FalkLanger Phys Rev E Stat Phys Plasmas Fluids Relat Interdiscip Topics 57(6):7192Ð7205
%MaloneyLemaitre Phys Rev E Stat Nonlin Soft Matter Phys 74(1):016118.
 This behaviour makes the measurements, and even the definition of elastic moduli quite involved. 
In a series of works Procaccia et al. have given evidences that in some models of glasses, such as Lennard-Jones mixtures (and variants), non-linear elastic moduli display diverging fluctuations and linear elastic moduli differ depending on the way they are defined from the stress-strain curve \cite {HKLP11,DPSS15}. Another independent research 
stream has focused on the understanding of the jamming and glass transitions of hard spheres both from 
real space and mean-field theory perspectives \cite{LNVW10,CKPUZ14NatComms}.
% AJ Liu, SR Nagel, W Van Saarloos, M Wyart "The jamming scenario-an introduction and outlook." "Heterogeneities in amorphous %systems under shear." %Dynamical %Heterogeneities in Glasses, Colloids, and Granular Media (2011).
The exact solution obtained in the limit of infinite dimensions revealed that by increasing the pressure a hard sphere glass displays 
a transition within the solid phase, where multiple arrangements emerge as different competing solid phases \cite{KPZ12,KPUZ13}. 
This is called Gardner transition in analogy with previous results in disordered spin models \cite{Ga85,GKS85}.
Recent simulations
have confirmed that in three dimensions these different arrangements become more and more long-lived, possibly leading to an ergodicity 
breaking \cite{BCJPSZ15}. These mean-field analysis complements and strengthens all the remarkable results found in the last two decades   
on jammed hard spheres glasses. The major outcome of these real space studies was the discovery that amorphous jammed solids 
are marginally stable, i.e. characterised by soft-modes and critical behaviour and in consequence by properties which are very different from the ones of usual crystalline solids \cite{OSLN03,WNW05,Wy12}.  
%LiuPointJ OÕHern, C. S., Silbert, L. E., Liu, A. J., & Nagel, S. R. (2003). Jamming at zero temperature and zero applied stress: The %epitome of disorder. Physical Review E, 68(1), 011306.
Within mean-field theory this is a consequence
of a more general marginal stability emerging at the Gardner transition \cite{CKPUZ14NatComms,FPUZ15}. \\
Here we show that also models of structural glasses display this transition when decreasing the temperature and that this drastically affects their elastic behaviour. In particular, we reveal that elastic anomalies, such as the ones found in zero temperature simulations, are a signature of this phase transition. In order to show the existence and the properties of the Gardner transition in structural glasses we focus on a system of soft elastic spheres, which has been studied recently in several numerical simulations and shown to behave as canonical glass-formers \cite{BW09,schmiedeberg2011mapping,berthier2012finite}.  The interaction potential between particles reads
\beq
\hat V_{\mathrm{HSS}}( r)=\frac{V_0}{2}\left(1-\frac{r}{\DD}\right)^2\theta\left(1-\frac{r}{\DD}\right)
\label{potenziale}
\eeq
where $r$ is the distance between particles, $V_0$ is the interaction strength
and $\DD$ the interaction range. We choose this model since in the limit $V_0\to \infty$, it maps on hard spheres with diameter $\DD$.
This enables us to make connections with previous results on jamming.\\
In this work we want to study the elastic properties of amorphous solids created by thermal quenches and also by compression. 
Theoretically, these solids are actually ultra-viscous liquids observed on time-scales on which flow is absent. From the energy landscape  perspective \cite{DS01}, these are systems unable to escape from a given metabasin within the experimental time-scale.   
The large dimensional limit ($d\rightarrow \infty$) is particularly useful to analyse these long-lived 
amorphous metastable states. Because the life-time of metastable states 
diverges exponentially with $d$, one does not have to develop a full dynamical 
treatment but can instead resort to a generalised thermodynamic framework able to capture the 
properties of metastable states \cite{Mo95}. 
What is generically considered a weakness of mean-field theory--the inability of describing activated dynamics in the
super-cooled regime--becomes here an advantage. 
In the infinite dimensional limit meta-basins become very long-lived below 
a well defined temperature $T_{MCT}$, corresponding to the Mode Coupling Transition (MCT). 
Although in three dimensions the increase of the life-time of meta-basins  is not as sharp below $T_{MCT}$ (MCT becomes
a cross-over), in the experimentally relevant regime we are interested in, amorphous solids do become well defined metastable states.
Indeed, for realistic quenches (0.1-100 K/min), super-cooled liquids fall out of equilibrium 
at a temperature $T_g$ well below $T_{MCT}$, and their properties do not change with time (except on very large 
times where ageing sets in); hence, the generalised thermodynamic framework \cite{Mo95} we use is particularly adapted.\\ 
In order to solve the model (\ref{potenziale}) in the $d\to \infty$ limit and to study the properties of the metastable amorphous solids 
we use the replica method, whose order parameter is the mean square displacement between couples of replicas ($x^{(a)}_i$ is the position of the particle $i$ in replica $a$):
\beq
\D_{ab}=\frac{d}{N\DD^2} \sum_{i=1}^N\left|x_i^{(a)}-x_i^{(b)}\right|^2
\eeq
Roughly speaking, this order parameter allows to study the statistical properties of the metabasins and probe their ruggedness: a breaking of replica symmetry means that that different replicas are trapped in different minima and, hence, that an ergodicity breaking transition
has taken place within the solid phase.
\begin{figure}
\centering
\includegraphics[scale=0.35]{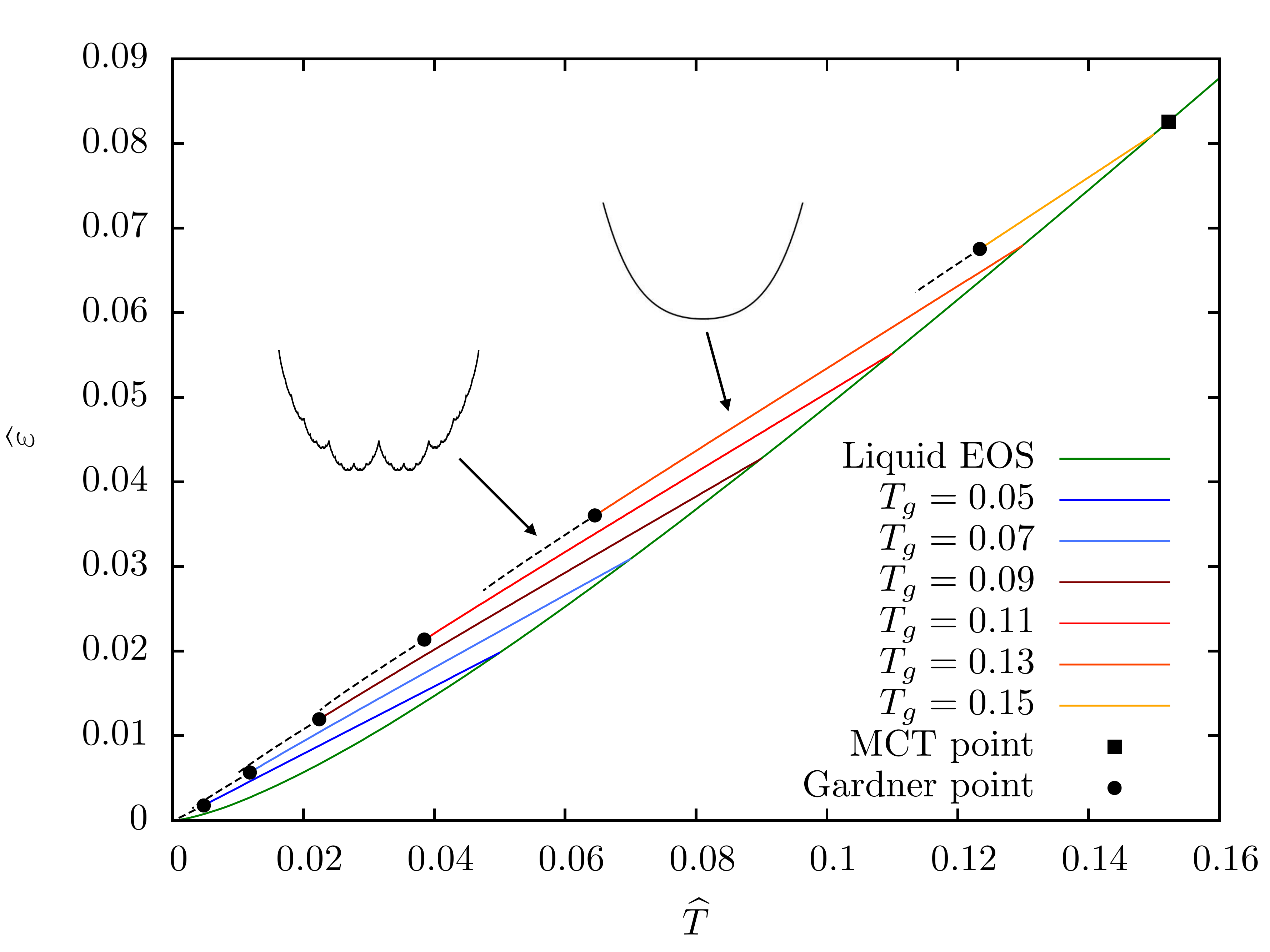}
\caption{Energy vs temperature (in rescaled units, see SM) for $\hat \varphi=\varphi 2^d/d=8$. 
EOS denotes the equilibrium line obtained by the equation of state. The other lines correspond to amorphous solids created by  
different cooling rate (slower to higher from bottom to top). Lines becomes dashed when the 
Gardner transition takes place. The change in the free-energy landscape at the Gardner transition is shown pictorially.}
\label{phasediag-fig}
\end{figure}  
This method has been developed and explained in full detail in several recent works on hard spheres \cite{CKPUZ14JSTAT,RUYZ15,RU15}. Hence, 
here we directly present our main results and refer to SM for the key technical steps (the whole derivation will be shown elsewhere \cite{BU16u}). \\
Henceforth we consider packing fractions such that $T_{MCT}(\phi)>0$
and focus on glass states formed by slow quenches below it ($T_{MCT}(\phi)$ raises from zero at a well defined 
packing fraction $\phi$ which in three dimensions should correspond to $\phi_{MCT}\simeq 0.58$).  
For a given cooling rate the system follows the equilibrium line in the energy-temperature plane until 
it falls out of equilibrium and becomes an amorphous solid at $T_g$. We have computed both the equilibrium and the amorphous solid
branches, as shown in Fig.1 for a given packing fraction. The main results is that generically by decreasing the temperature amorphous solids  undergo a Gardner transition at a temperature $T_G(\phi)$. The lower is the glass transition temperature $T_g$ the 
more one has to cool in order to reach $T_G(\phi)$.   
\begin{figure}
\centering
\includegraphics[scale=1]{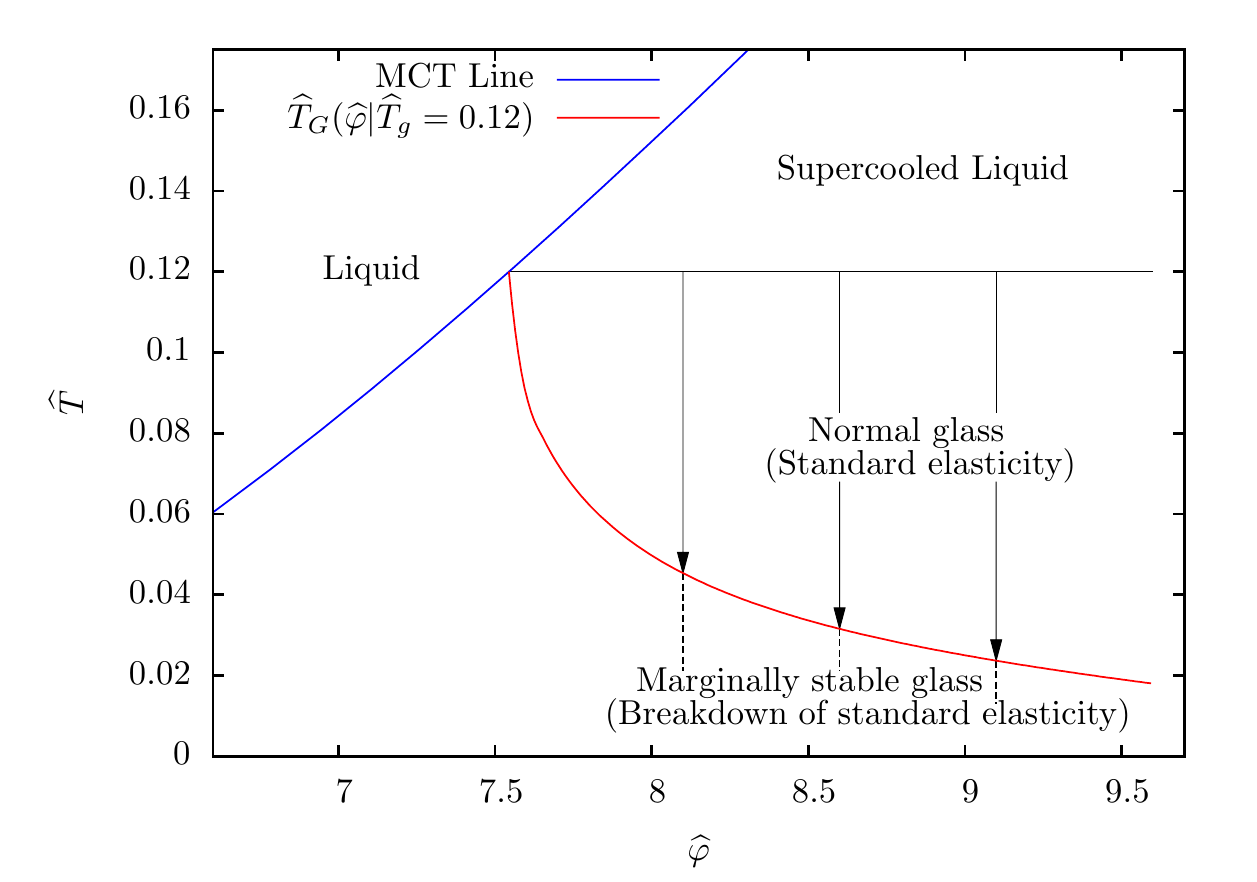}
\caption{Evolution of the Gardner transition temperature (red line) obtained by cooling amorphous solids at fixed $\varphi$. All  amorphous solids are formed at the same $\widehat T_g=0.12$. The blue line denotes the MCT transition temperature as a function of $\widehat \varphi$.}
\label{phasediag-fig}
\end{figure}  
By comparing the results obtained for amorphous solids formed at the same $T_g$ at different packing factions, we find 
that $T_G(\varphi)$ decreases when $\varphi$ is augmented, as shown in Fig. 2 (for too small densities, when $T_{g}$ crosses $T_{MCT}$, it is simply not possible to create a solid).    
We conclude this analysis by considering glasses formed by compression 
to make a relationship with studies on hard spheres and jamming. In this case, we find  
first a direct and then an inverse Gardner transition as shown in Fig. 3. 
The results obtained for glasses formed at the same $\phi_g$ at different temperatures show that the higher is the temperature the smaller is the extent of the Gardner phase, thus creating a dome in the $T-\varphi$ plane. 
This re-entrant behaviour is in agreement with 
very recent studies on the spectrum of harmonic vibrations of elastic sphere glasses \cite{FPUZ15, CCPPZ15}. \\
In summary, we have shown that amorphous solids undergo by cooling the same transition toward a marginal glass state found 
for hard spheres \cite{KPUZ13}. 
%We have also discussed its evolution at different densities and using protocols, such as compression, relevant for colloidal glasses. 
\begin{figure}
\centering
\includegraphics[scale=0.35]{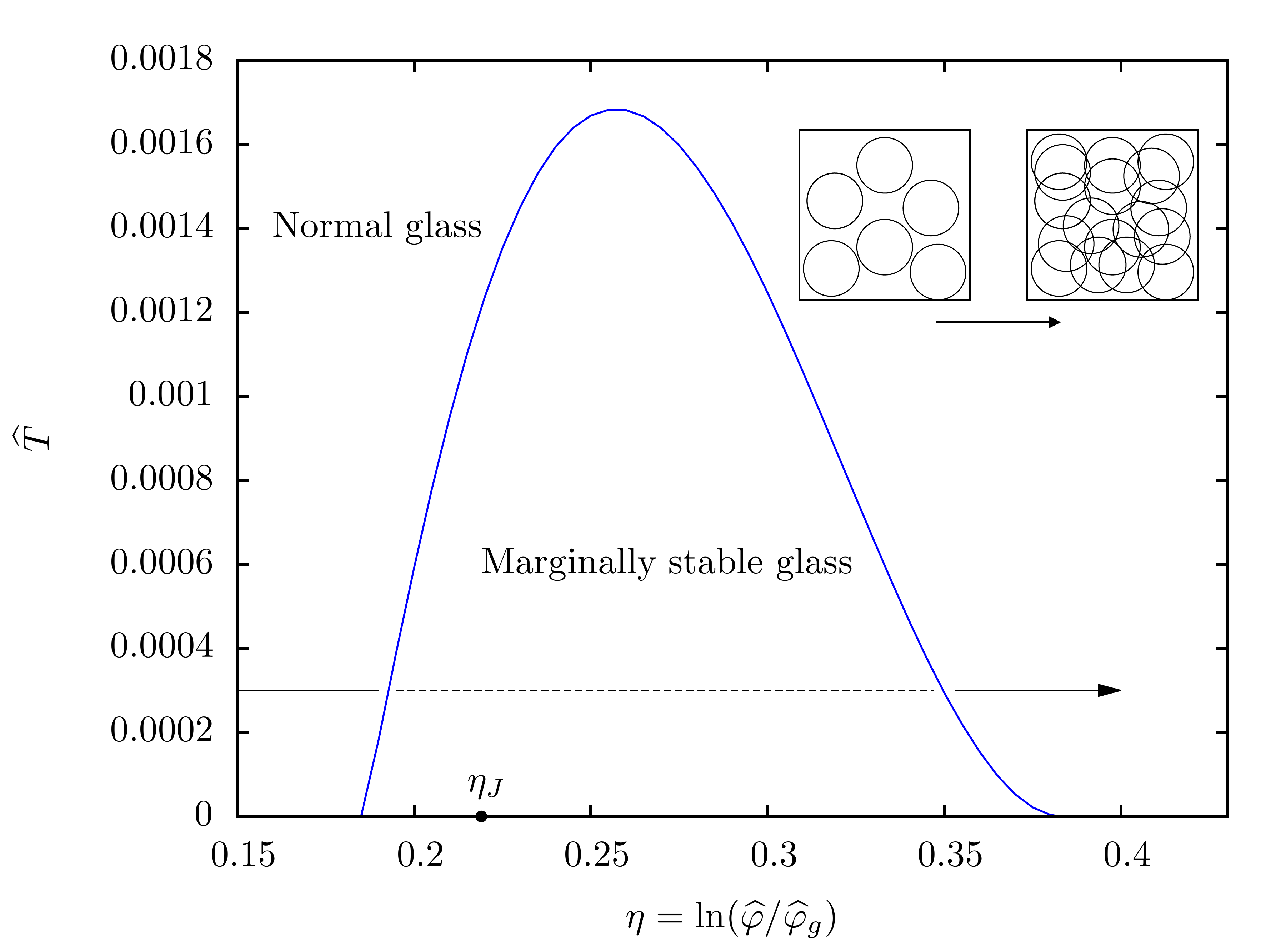}
\caption{Gardner critical packing fractions $\phi_G$ obtained by compressing amorphous solids at fixed temperature. All  amorphous solids are formed at the same $\widehat \varphi_g=8$. The location of the jamming point is denoted by $\eta_J$.}
\label{phasediag-fig}
\end{figure}  
We can now turn to our main concern which is the change in the elastic properties of the solid approaching the Gardner transition. For a normal elastic solid, e.g. a crystal, a small shear strain $\gamma$ induces a change in free energy per unit volume equal to
\beq
\begin{split}
\frac{\mathcal{F}_{el}}{V}=
\frac{\mu_2}{2}\g^2+\frac{\m_4}{4!}\g^4+É\ldots
\label{general_expansion}
\end{split}
\eeq
where $V$ is the volume and $\m_n$ is the $n$-th order elastic modulus: $\m_n=\frac{d\sigma^{n-1}}{d^{n-1}\gamma}$ where $\sigma$ is the stress ($\m_2$ is the usual linear shear modulus). This is also true for amorphous solids {\it but only above the Gardner transition}. 
Our explicit computation in the limit of infinite dimensions shows that in this regime all elastic moduli are well defined (up to fluctuations of the order $1/\sqrt{V}$) and that they depend on the 
glass state only through the value of $T_g$, i.e. the speed of the quench used to form the glass, and of the applied temperature and 
packing fraction \footnote{We are working in the NVT ensemble but of course our results can be translated to other cases, for example the more experimentally relevant NPT one.}. The situation changes drastically approaching the Gardner transition line at which the fluctuations of all elastic moduli blow up. Although averages remain featureless, the 
fluctuations from one glass state to another, rescaled by their typical value $1/\sqrt{V}$, diverge as:
\beq\label{div}
\overline{(\delta \mu_n \sqrt{V})^2} \sim \frac{1}{(T-T_G)^{2n-3}}\qquad , \qquad  \overline{(\delta \mu_n \sqrt{V})^2} \sim \frac{1}{|\phi-\phi_G|^{2n-3}}
\eeq
where the right and left expressions correspond to different protocols to induce the transition (cooling and compression).
This increase leads to giant fluctuations at the Gardner transition. Finite size (mean-field) scaling implies that at the transition 
$\overline{(\delta \mu_n \sqrt{V})^2}\sim V^{2n/3-1}$. For $n=2$ fluctuations are subleading, hence the linear elastic shear modulus is regular and well defined: $\mu_2\simeq \overline \mu_2 +O(V^{-1/6})$. Instead all non-linear moduli are not: their fluctuations diverge as $V^{n/3-1}$ and completely overwhelm the average, which remains finite but is not representative of the typical behaviour. Note, moreover, that all odds moduli, that vanish by symmetry in ordered solids, can be neglected above $T_G$ only. At $T_G$ they also blow up and instead of being of the order of $1/\sqrt{V}$ (and zero in average) they diverge as $V^{n/3-1}$ and fluctuate. All these results signal that standard elastic behaviour breaks down at $T_G$. Below $T_G$,
even an infinitesimal deformation leads in the thermodynamic limit to ageing and time-dependent shear moduli. In this regime, elastic moduli depend
on the history and on the protocol used to measure them. Only strains whose amplitude scales to zero with $V$ do not lead to ageing and irreversible behaviour \cite{DPSS15}. The elastic moduli computed in this way, called quenched  in \cite {DPSS15} and zero-field cooled in \cite{NYZ15}, are a property of the meta-basin to which the system belongs. They are characterised by the same divergent fluctuations found at $T_G$. This is a consequence of the marginal stability of glasses within the whole Gardner phase.  
We derived our results in a specific realistic model in the limit of infinite dimensions. However, our findings 
go beyond the specific $d\to \infty$ computation we presented. Indeed one can obtain them using a Landau theory as shown in the supplementary material. Similarly to the existence of diverging magnetic responses at a ferromagnetic transition, the breakdown of elastic behaviour and the divergence of non-linear elastic moduli are the generic signature of the Gardner transition \footnote{As a matter of fact, 
our results also hold mutatis mutandis for spin-glasses in a field and provide new ways to test for the existence of the Almeida-Thouless line \cite{BU16u}.}.\\  
In conclusion, our results unveil that the jerky elastic behaviour displayed by amorphous solids at low temperature could be related to the existence of a phase transition within the glass phase. Our exact solution in the limit of infinite dimensions
characterises the dependence of this transition on the control parameters ($T$, $\varphi$) and protocols (cooling, compression), providing guidance for experimental tests both in structural, colloidal and possibly granular glasses.
In order to substantiate our predictions, it is impelling to extensively study by experiments and by simulations whether standard elastic behaviour breaks down at a well defined temperature in the 
way identified in this work. The simulations of model systems quenched at zero temperature by Procaccia et al. display very promising results: the linear elastic intra-state modulus is found to be well defined, whereas $\mu_3$ has fluctuations of $O(1)$ and $\mu_4$ shows diverging fluctuations as we found. On the theoretical side, it is important to go beyond the Landau theory in order to obtain more quantitative predictions on the value of the critical exponents controlling the divergence of the elastic moduli. 
All that opens the way toward new
research directions aimed at revealing the true nature of glasses. As suggested by several recent research results on jamming and amorphous plasticity, glasses might not be just liquids having stopped to flow but an entirely different new kind of solids \cite{GLN14,Bi14,CKPUZ14NatComms}.

 %that for finite $V$ $\overline{(\delta \mu_n \sqrt{V})^2} \sim \frac{1}{\epsilon^{n/2}} f_{FSS}^{(n)}(V\epsilon^{1/3})
%$ \footnote{The scaling variable controlling finite size scaling is $V\epsilon^{1/3}$ for FRSB transitions, and it can be explicitly verified for the Gardner transition from the field theoretical analysis \cite{UB}.}. The scaling function $f_{FSS}^{(n)}(V\epsilon^{1/3})$ is such that $f_{FSS}^{(n)}(x)\rightarrow 1$ for $x\rightarrow \infty$  in order to recover eq. (\ref{div}) and 
%$f_{FSS}^{(n)}(x)\sim x^{3n/2}$ for $x\rightarrow 0$ in order for the $\epsilon$ dependence to drop out in the $\epsilon \rightarrow 0$ limit.

\emph{Acknowledgments} 
We thank J.-P. Bouchaud, P. Charbonneau, S. Franz, I. Procaccia, G. Tarjus, F. Zamponi for useful discussions.
We acknowledge financial support from the ERC grant NPRGGLASS.

%\bibliographystyle{mioaps}
%\bibliography{HS}

\clearpage

\centerline{ \bf Supplementary material}

\centerline{\bf Breakdown of elasticity in amorphous solids}

\medskip

\centerline{Giulio Biroli and Pierfrancesco Urbani}

\section{Introduction} 
We consider a system of spheres interacting through a central potential $\hat V(r)$. A typical example is the Harmonic Soft Sphere interaction potential
\beq
\hat V_{\textrm{HSS}}(r)=\frac{V_0}{2}\left(1-\frac{|r|}{\DD}\right)^2\theta\left(1-\frac{|r|}{\DD}\right)
\label{potenziale}
\eeq
where $\DD$ is the diameter of the sphere that fixes the interaction range and $V_0$ is the interaction strength. Taking the limit $V_0\to \infty$ one gets back the Hard Sphere model.
%In a mean field setting, like for example in the infinite dimensional limit, we can carefully define a metastable glassy state. This can be fixed in two different ways: either by fixing the amorphous boundary conditions to the system \cite{BB04} or by constraining the configurations of the system to be close to an amorphous lattice that represents the amorphous state \cite{FP95}. 
A glassy state $\a$ can be characterized by few control parameters such as the packing fraction $\f$ and the inverse temperature $\b$ 
at which it is created, i.e. at which the super-cooled liquid fell out of equilibrium.  We shall denote it $\a(\f, \b)$.
It can be thought as a metabasin of configuration of phase space that is explored ergodically for timescales shorter than the $\a$-relaxation time. More precisely, we shall say that we \emph{prepare} the system in a glass state $\a(\varphi_g, \beta_g)$ when the configurations that are sampled by the dynamics are equilibrium configurations of the metabasin to which the state $\a$ belongs to.
Once a glassy state is prepared at a given point $(\f_g, \b_g)$, we can look at how it changes when the control parameters are changed towards another state point $(\f,\b)$. In particular we can compute the free energy of the glassy state once \emph{followed} to this new state point and it is given by
\beq
f\left[\a(\f_g,\b_g), \b, \f\right]=-\frac{1}{\b V} \ln \int_{X\in \a(\f_g, \b_g)} \de X \eee^{-\b \VV[X; \f]}
\label{Free_energy_general}
\eeq
where $\VV[X;\f]=\sum_{i<j}\hat V(|x_i-x_j|)$. The sum over the configurations $X$ is done in such a way that they all belong to the same ergodic component characterizing the glass state $\a(\widehat \varphi_g,\widehat \b_g)$.
The packing fraction can be changed by changing the diameter of the spheres in (\ref{potenziale}), \cite{RUYZ15_bis}. 
%The calculation of the free energy (\ref{Free_energy_general}) can be performed using the Franz-Parisi potential \cite{FP95} and it has been done for Hard Spheres in \cite{RUYZ15_bis}.
In order to compute the elastic response of the system we need to couple it to an external strain $\g$. In this way the interaction potential is changed due to the change in the shape of the box in which the system is placed. We denote the interaction potential in presence of the strain as $\VV_\g[X; \f]$.
If the shear strain is small, the change in the free energy (\ref{Free_energy_general}) is given by
\beq
\begin{split}
V\left(f\left[\a(\f_g,\b_g), \b, \f; \g\right]-f\left[\a(\f_g,\b_g), \b, \f; 0\right]\right)&=V \s \g + V\frac{\mu_2}{2}\g^2+V \frac{\mu_3}{3!}\g^3+V\frac{\m_4}{4!}\g^4+\ldots
\label{general_expansion}
\end{split}
\eeq
Since the amorphous state is a random structure we expect the elastic coefficients $\s, \mu_2, \mu_3$ and $\mu_4$ to be random variables characterized by a proper probability distribution.
From the expansion (\ref{general_expansion}) it is clear that the shear stress is given by $\de f\left[\a(\f_g,\b_g), \b, \f; \g=0\right]/\de \g=\s$ while the shear modulus is $\mu_2$. 
The scaling in $V$ in (\ref{general_expansion}) can be deduced from extensivity.
%and from the fact that the ergodic component $\a(\f_g,\b_g)$, once followed at $(\f,\b)$ does not cross any phase transition up to the Gardner point. 
Moreover we expect that once we average the free energy over all glassy states $\a(\f_g, \b_g)$, we get a function that is symmetric under $\g\to -\g$. This implies that
\beq
\overline{\s}=0\ \ \ \ \ \ \ \overline{\mu_3}=0
\eeq 
where we have denoted with an overline the average over different glassy states $\a(\varphi_g,\b_g)$.
Conversely, the first moment of $\mu_2$ and $\mu_4$ is expected to be different from zero.
If the glassy state $\a(\f_g,\b_g)$ is followed up to a point that is close to a second order phase transition, like the Gardner point \cite{KPUZ13_bis}, we expect a dramatic change in the probability distribution of the elastic moduli, due to the presence of the soft modes developed at the transition.

\section{Free replica sum expansion}
Since we want to average over all glassy states $\a$ that can be found at $(\f_g,\b_g)$ we need to introduce replicas to handle the logarithm that appears in (\ref{Free_energy_general}).
In order to compute all the different cumulants it is very useful to consider each replica being subjected to a different strain $\g$.
In this way we get $s$ systems, each one subjected to a different shear strain $\g_a$. We thus define a replicated free energy
\beq
\begin{split}
&W[\f, \b|\f_g,\b_g;\{\g_a\}]=-\frac{1}{\b V}\ln\overline{\prod_{a=1}^s\int_{X^{(a)}\in\a(\f_g,\b_g)}\de X^{(a)}\eee^{-\b \VV_{\g_a}[X^{(a)}; \f]}}\:.
\end{split}
\label{replicated_free_energy}
\eeq
Once the average over the glassy states $\a(\varphi_g,\b_g)$ is taken, we end up with a replicated system of $s+1$ replicas \cite{FP95, BFP97, RUYZ15_bis} the first one being a representative configuration of the glass state planted at $(\varphi_g, \beta_g)$; at the end we will consider the limit $s\to 0$. Note that the initial glass state at $(\varphi_g,\beta_g)$ is unstrained so that $\g_1 =0$.

If we expand this function aroung $\{\g_a=0\}$ we get a free replica sum expansion \cite{LMW08, TT04}
\begin{widetext}
\beq
\begin{split}
&W[\f, \b|\f_g,\b_g;\{\g_a\}]-W[\f, \b|\f_g,\b_g;\{\g_a=0\}]=\frac{\overline \mu_2}{2}\sum_{a=2}^{s+1}\g_a^2-\frac{V\overline{\s^2}^c}{2}\left(\sum_{a=2}^{s+1}\g_a\right)^2\\
&-\frac{V^3\overline{\sigma^4 }^c}{4!}\left(\sum_{a=2}^{s+1}\g_a\right)^4
-\frac{V\overline{\s\mu_3}^c}{3!}\left(\sum_{a=2}^{s+1}\g_a\right)\left(\sum_{a=2}^{s+1}\g_a^3\right)-\frac18V \overline{\mu_2^2}^c\left(\sum_{a=2}^{s+1}\g_a^2\right)^2\\
&+\overline{\s^2\mu_2 V^2}^c\left(\sum_{a=2}^{s+1}\g_a\right)^2\left(\sum_{a=2}^{s+1}\g_a^2\right)+\frac{\overline \mu_4}{4!}\left(\sum_{a=2}^{s+1}\g_a^4\right)+\dots\:.
\end{split}
\label{cumulant_expansion}
\eeq
\end{widetext}
Due to extensivity, we have that $\overline{\s^2}^c\sim 1/V$, $\overline{\s^4}^c\sim 1/V^3$, $\overline{\mu_2^2}^c\sim 1/V$, $\overline{\s \mu_3}^c\sim 1/{V}$ and $\overline{\s^2 \mu_2}^c\sim 1/{V^2}$.
Moreover, due to the symmetry $\{\g_a\to -\g_a\}$ we must have $\overline{\s\mu_2}^c=0$.

The replicated free energy (\ref{replicated_free_energy}) can be computed exactly in the infinite dimensional limit \cite{CKPUZ14JSTAT_bis, RUYZ15_bis} where we can carefully define metastable glassy states since nucleation and non-perturbative effects are highly suppressed.
%, a situation that physically occurs in strongly equilibrated glasses. 
This is given in terms of an order parameter that is nothing but the distance between different replicas. If we denote as $x^{(a)}_i$ the position of the sphere $i$ in replica $a$, we can define the mean square displacement (MSD) between couples of replicas as
\beq
\D_{ab}=\frac{d}{N\DD^2} \sum_{i=1}^N\left|x_i^{(a)}-x_i^{(b)}\right|^2
\eeq
where $d$ is the spatial dimension and it has been added in order to have a finite MSD matrix in the large dimensional limit.
The matrix $\Delta_{ab}$ must be fixed by a saddle point equation $\partial W/\partial \Delta_{ab}=0$. This means that the calculation of the cumulants in (\ref{cumulant_expansion}) can be done by expanding in powers of $\{\g_a\}$ the saddle point replicated free energy $W$.   
This is given by
\beq
\begin{split}
&W[\f, \b|\f_g,\b_g;\{\g_a\}]-W[\f, \b|\f_g,\b_g;\{\g_a=0\}]=\frac 12 \sum_{a,b=2}^{s+1} \mu_{ab}\g_a\g_b+\frac{1}{4!}\sum_{a,b,c,d=2}^{s+1}\chi_{abcd}\g_a\g_b\g_c\g_d+\ldots
\end{split}
\eeq
where
\beq
\mu_{ab}=\frac{\de W}{\de \g_a\de \g_b}=\frac{\partial W}{\partial \g_a\partial \g_b}
\label{derivata_seconda}
\eeq
\beq
\begin{split}
\chi_{abcd}=\left.\frac{\de^4 W}{\de \g_a\de \g_b\de \g_c\de \g_d}\right|_{\{\g_a=0\}}&=-\sum_{\a\neq \b}\sum_{\mu\neq \nu}[M^{-1}]_{\a\b;\mu\nu}\left[V_{\a\b}^{ab}V_{\mu\nu}^{cd}+V_{\a\b}^{ac}V_{\mu\nu}^{bd}+V_{\a\b}^{ad}V_{\mu\nu}^{bc}\right]\\
&+\left.\frac{\partial^4 W}{\partial \g_a\partial \g_b\partial \g_c\partial \g_d}\right|_{\{\g_a=0\}}
\label{derivata_quarta}
\end{split}
\eeq
and where
\beq
V_{\a\b}^{ab}=\left.\frac{\partial^3 W}{\partial \g_a\partial \g_b\partial \D_{\a\b}}\right|_{\{\g_a=0\}}
\eeq
\beq
M_{\a\b;\mu\nu}=\left.\frac{\partial^2 W}{\partial \Delta_{\a\b}\partial \Delta_{\mu\nu}}\right|_{\{\g_a=0\}}\:.
\eeq
All the derivative are computed setting the order parameter $\Delta_{ab}$ to its saddle point value at $\{\g_a=0\}$.
From Eq.s (\ref{derivata_seconda}-\ref{derivata_quarta}) we see that the coefficients $\mu_{ab}$ are always regular while the quartic coefficients $\chi_{abcd}$ can develop divergencies due to the fact that the operator $M_{\a\b;\mu \nu}$ can develop zero modes. This happens at the Gardner transition point \cite{KPUZ13_bis}. The Gardner instability has been found in the context of the Hard Sphere model in \cite{KPUZ13_bis, RUYZ15_bis} and in the next section we will generalize these results to thermal systems. What we have done up to now is nothing but a Landau theory for the Gardner transition point.

In the normal glass phase, before reaching the Gardner instability, the saddle point MSD matrix has a 1-step replica symmetry breaking (1RSB) fashion (see the next section) and thus
the form of the tensors $M$ and $V$ is simple. The tensor $M$ has been discussed in \cite{KPUZ13_bis,RUYZ15_bis}
while $V$ is given by
\beq
\begin{split}
&V_{\a\b}^{ab}=\delta_{ab}\left[\tilde G \Omega_{\a\b}^a+\tilde H\Gamma^a_{\a\b}\right]+(1-\d_{ab})\left[\tilde PT^1_{ab;\a\b}+\tilde QT^2_{ab;\a\b}+\tilde RT^3_{ab;\a\b}\right]
\end{split}
\eeq
where the tensors $\Omega,\  \G,\ T^1,\ T^2$ and $T^3$ are defined by
\beq
\begin{split}
&\Omega_{\a\b}^a=\frac{\delta_{a\a}+\delta_{a\b}}{2}\ \ \ \ \G_{\a\b}^a=1\ \ \ \ T^1_{ab;\a\b}=\frac{\delta_{a\a}\delta_{b\b}+\delta_{a\b}\delta_{b\a}}{2}\\
&T^2_{ab;\a\b}=\frac{\delta_{a\a}+\delta_{b\b}+\delta_{a\b}+\delta_{b\a}}{4}\ \ \ \ T^3_{ab;\a\b}=1\:.
\end{split}
\eeq
At the Gardner transition point the operator $M$ develops a zero eigenvalue, the replicon, that we denote by $\l_R$. The corresponding eigenvector is a matrix $\delta_R\D_{ab}$ that satisfies the conditions
\beq
\begin{split}
&\sum_{a(\neq b)}\delta_R\Delta_{ab}=\sum_{b(\neq a)}\delta_R\Delta_{ab}=0\ \ \ \ \ \ \ \ a,b=2,\ldots s+1\\
&\delta_R \Delta_{1a}=0 \ \ \ \ \ \forall a=2,s+1\:.
\end{split}
\eeq
This means that the first term on the right hand side of (\ref{derivata_quarta}) can be divergent due to the zero mode. In order to evaluate the divergent parts of the cumulants in (\ref{cumulant_expansion}) we define the projector on the replicon subspace
\beq
\begin{split}
P^\parallel_{ab;cd}=\frac 12 \left(T^1_{ab;cd}+ T^2_{ab;cd}+T^3_{ab;cd}\right) \ \ \ \ \ \ a,b,c,d=2,\ldots s+1
\end{split}
\eeq
and it is zero whenever one of the replica index is equal to 1.
From the quartic derivatives we get that the singular part of the quartic derivative of the replicated free energy is given by
\beq
\begin{split}
&\left.\frac{\de^4 W}{\de \g_a\de \g_b\de \g_c\de \g_d}\right|_{\textrm{singular}}=-\frac{1}{\l_R}\sum_{(\a\neq \b)(\mu\neq \nu)}^{s}P^\parallel_{\a\b;\mu\nu}\left[V_{\a\b}^{ab}V_{\mu\nu}^{cd}+V_{\a\b}^{ac}V_{\mu\nu}^{bd}+V_{\a\b}^{ad}V_{\mu\nu}^{bc}\right]
\end{split}
\eeq
From this expression we finally get that the \emph{divergent} part of the quartic cumulants of (\ref{cumulant_expansion}) is given by
\beq
\begin{split}
V\overline{\s \mu_3 }^c&=-\frac{3\tilde P^2}{2\l_R}\\
V\overline{\mu_2^2}^c&=\frac{3\tilde P^2}{2 \l_R}\\
V^3\overline{\s^4}^c&=\frac{3\tilde P^2}{2\l_R}\\
\end{split}
\eeq
The averages $\overline{\mu_4}$ and $V^2\overline{\s^2 \mu_2}^c$ have no diverging contribution, i.e. they are finite at the transition. 
This is true for all $\overline{\mu_n}$. However, by going to higher orders in the free replica sum expansion we can get higher order correlation functions of the elastic moduli and in particular we can obtain the variances of the non-linear elastic moduli. 
These diverge as discussed in the text. The average finite values $\overline{\mu_n}$ are therefore not representative of the typical behaviour, which is dominated by the diverging fluctuations. As written in the main text, we have verified that these divergences, 
due to the vanishing of the replicon eigenvalue, also hold in the low temperature phase below $T_G$.  
All the detailed calculations will be presented elsewhere \cite{BU16u_bis}.

\section{The Gardner transition in thermal systems}
The Gardner transition that is responsible for the breakdown of the theory of elasticity in amorphous solids has been firstly detected in \cite{KPUZ13_bis, RUYZ15_bis} in the case of the Hard Sphere model. Here we want to generalize the theory to thermal systems. Let us consider a system of spheres that interact through an interaction potential of the following kind
\beq
\hat V(r)=\tilde v\left(d\left(\frac{|r|}{\DD}-1\right)\right)
\eeq
where $\tilde v$ is a generic potential that decreases sufficiently fast at infinity \cite{KMZ15}.
The Harmonic Soft Sphere potential is exactly of this form. Indeed, in that case we have (we set $V_0=1$ without loss of generality)
\beq
\tilde v_{\mathrm{HSS}}(h)=\frac{h^2}{2 d^2}\theta(-h)=\frac{1}{d^2}v_{\mathrm{HSS}}(h)\ \ \ \ \ \ \ \ v_{\textrm{HSS}}(h)=\frac{h^2}{2}\theta(-h)\:.
\eeq
It is convenient to define a reduced temperature $\wh \b=\b/d^2$ so that the Boltzmann factor becomes
\beq
\eee^{-\b \hat V_{\textrm{HSS}}(r)} = \eee^{-\wh \b v_{\mathrm{HSS}}(h)}\:.
\eeq
For a generic potential we will define $v(h)$ such that
\beq
\eee^{-\b \hat V(r)} = \eee^{-\wh \b v(h)}
\eeq
where $\widehat \beta$ is an inverse temperature properly rescaled with the dimension.

We now consider the limit $d\to \infty$.
In order to do that we need to consider scaled control parameters that are $\widehat \b$ and $\widehat \varphi = 2^d \varphi/d$.
By extending the calculations of \cite{RUYZ15_bis, RU15_bis} we obtained the expression of the free energy of a glass state prepared at $(\widehat \varphi_g, \widehat \beta_g)$ and followed up to $(\widehat \varphi=\widehat\varphi_g \eee^\eta, \widehat \beta)$.
Within a fullRSB ansatz \cite{CKPUZ14JSTAT_bis, RU15_bis}, the order parameter $\Delta_{ab}$ is given by
\beq
\begin{split}
\Delta_{1a}&=\Delta^r \ \ \ \ \ \forall a=2,\ldots,s+1\\
\Delta_{ab}&\to \{0,\Delta(x)\} \ \ \ \ \forall a,b=2,\ldots,s+1\ \ \ \ \ \ x\in[0,1]
\end{split}
\eeq
and the free energy of the glass state prepared at $(\widehat \varphi_g, \widehat \beta_g)$ and followed up to $(\widehat \varphi=\widehat\varphi_g \eee^\eta, \widehat \beta)$ is given in by
\beq
\begin{split}
-\beta f\left[\a(\f_g,\b_g), \b, \f\right] =\ &\frac{d}{2} + \frac{d}{2}\log\left(\frac{\pi\la\D\ra}{d^2}\right) - \frac{d}{2}\int_0^1\frac{\de y}{y^2}\log\left(\frac{\la\D\ra + [\D](y)}{\la\D\ra}\right) + \frac{d}{2}\frac{\D_R}{\la\D\ra}\\ 
&+\frac{d\wh\varphi_g}{2} \int_{-\infty}^\infty \de h \, \eee^{h} g_{\D_R}(1,h+\D_R/2)f(0, h-\eta +\D(0)/2,\widehat \beta).
\end{split}
\label{eq:sfull}
\eeq
where
\beq
\langle \Delta \rangle=\int_0^1 \de x \Delta(x) \ \ \ \ \ \ [\Delta](x)= x\Delta(x) - \int_0^x \de y\Delta(y)
\eeq
and where $\Delta_R=2\Delta^r - \Delta(0)$.
Moreover
\beq
g_\L(1,x;\widehat \b)=\int_{-\infty}^\infty \frac{\de y}{\sqrt{2\pi \L}} \exp\left[-\frac{y^2}{2\L }-\widehat \beta v(x-y)\right]
\eeq
while the function $f$ together with $\Delta(x)$ and $\Delta^r$ satisfy the following saddle point equations
\beq
\begin{split}
f(1,h, \widehat \beta)&=g_{\Delta(1)}(1,h,\widehat \beta)\\
\frac{\partial f}{\partial x}&=\frac 12\frac{\dot G(x)}{x}\left[\frac{\partial^2 f}{\partial h^2}+x\left(\frac{\partial f}{\partial h}\right)^2\right]\\
P(0,h)&=\eee^{h+\eta-\Delta(0)/2}g_{\Delta_R}\left(1,h+\eta-\frac{\Delta(0)+\Delta_R}{2},\widehat \beta_g\right)\\
\dot{P}(x,h)&= -\frac{\dot{G}(x)}{2x}\left[P''(x,h)-2x(P(x,h)f'(x,h))'\right]\\
\frac{1}{G(0)}&=-\frac{\wh \f_g}{2}\int_{-\infty}^\infty \de h P(0,h)\left(f''(0,h)+f'(0,h)\right)\\
\frac{\Delta_R}{G(0)^2}&=\frac{\wh \f_g}{2}\int_{-\infty}^\infty\de h\, P(0,h) \left(f'(0,h)\right)^2\\
\k(x)&=\frac{\wh \f_g}{2}\int_{-\infty}^\infty \de h P(x,h)\left(f'(x,h)\right)^2\\
\frac{1}{G(x)}&=\frac{1}{G(0)} +x\k(x)-\int_{0}^x\de y\k(y)\ \ \ \ \ \ \ \ \ \ \ \ x>0\\
G(x)&=x\Delta(x)+\int_{x}^1\de z\Delta(z)\ \ \ \ \      \Longleftrightarrow\ \ \ \ \       \Delta(x)=\frac{G(x)}{x}-\int_x^1\frac{\de z}{z^2}G(z)\:.
\label{fullRSB_continue}
\end{split}
\eeq
Before reaching the Gardner transition point, in the normal glass phase where standard elasticity holds, these equations are solved by a 1RSB ansatz that corresponds to putting $\Delta(x)=\Delta$ where $\Delta$ satisfies the much simpler set of equations
\beq
\boxed{
\begin{split}
&\frac{2\Delta^r}{\Delta^2}-\frac{1}{\Delta}=\wh\f_g\int_{-\infty}^\infty  \de h\, \eee^{h}\frac{\partial}{\partial \Delta} \left[g_{2\Delta^r-\Delta}\left(1,h+\Delta^r-\Delta/2; \wh\b_g\right) \log g_{\Delta}\left(1, h -\eta +\Delta/2;\wh \b \right)  \right]\\
0&=\frac{2}{\Delta}+\wh \f_g\int_{-\infty}^\infty  \de h\, \eee^{h}\left[\frac{\partial}{\partial \Delta^r} g_{2\Delta^r-\Delta}\left(1,h+\Delta^r-\Delta/2; \wh\b_g\right)\right] \log g_{\Delta}\left(1, h-\eta +\Delta/2;\wh \b \right)
\end{split}}
\label{SP1RSB}
\eeq
The Gardner transition corresponds to the point where
\beq
\boxed{
0=-1+\frac{\widehat \varphi_g}{2}\Delta^2 \int_{-\infty}^\infty \de h \eee^{h+\eta-\frac{\Delta}{2}}g_{\Delta_R}\left(1,h+\eta-\frac{\Delta(0)+\Delta_R}{2},\widehat \beta_g\right)\left[\frac{\partial^2}{\partial h^2} \ln g_{\Delta}(1,h,\widehat \beta)\right]^2}
\label{replicone}
\eeq
where the replicon eigenvalue vanishes.
The rescaled energy of the system (in the infinite dimensional limit) at the 1RSB level is given by
\beq
\widehat \varepsilon=-\frac{\widehat \varphi_g}{2}\int_{-\infty}^\infty \de h \eee^h g_{\Delta_R}\left(1,h+\frac{\Delta_R}{2},\widehat \beta_g\right)\frac{\partial }{\partial \widehat \beta}\ln g_{\Delta}\left(1, h-\eta+\frac{\Delta}{2},\widehat \beta\right)
\label{E1RSB}
\eeq
Formulas (\ref{SP1RSB}), (\ref{replicone}) and (\ref{E1RSB}) have been solved and computed numerically to obtain the phase diagrams of the main text. Their complete derivation with an extension of the theory with shear strain will be given elsewhere \cite{BU16u_bis}.

\bibliographystyle{mioaps}

\end{document}